\documentclass[10pt,letterpaper]{article}
\usepackage{opex3}
\usepackage{color}
\usepackage{cite}

\begin{document}

\title{Radiative thermal rectification between SiC and SiO$_2$}
\author{Karl Joulain,$^{1,*}$ Youn\`es Ezzahri,$^1$ J\'er\'emie Drevillon,$^1$ Beno\^\i t Rousseau,$^2$ and Domingos De Sousa Meneses,$^{3,*}$}
\address{$^1$ Institut Pprime, Universit\'e de Poitiers-CNRS-ENSMA, TSA 41105,  F-86073 Poitiers, France\\ 
$^2$LTN, CNRS UMR 6607,  B.P 90604, F-44306 Nantes,  France\\
$^3$ CNRS, CEMHTI UPR3079, Universit\'e d'Orl\'eans, F-45071 Orl\'eans, France}

\email{$^*$karl.joulain@univ-poitiers.fr, desousa@cnrs-orleans.fr}

\date{\today}
\begin{abstract}
By means of fluctuational electrodynamics, we calculate radiative heat flux between two planar materials respectively made of SiC and SiO$_2$. More specifically, we focus on a first (direct) situation where one of the two materials (for example SiC) is at ambient temperature whereas the second material is at a higher one, then we study a second (reverse) situation where the material temperatures are inverted. When the two fluxes corresponding to the two situations are different, the materials are said to exhibit thermal rectification, a property with potential applications in thermal regulation. Rectification variations with temperature and separation distance are reported here. Calculations are performed using material optical data experimentally determined by Fourier transform emission spectrometry of heated materials between ambient temperature (around 300 K) and 1480 K.
It is shown that rectification is much more important in the near-field domain, i.e. at separation distances smaller than the thermal wavelength.  In addition, we see that the larger is the temperature difference, the larger is rectification. Large rectification is finally interpreted due to a weakening of the SiC surface polariton when temperature increases, a weakening which affects much less SiO$_2$ resonances. 
\end{abstract}

\bibliographystyle{osajnl}

%\bibliography{biblio_all.bib}

\section{Introduction}

Since the seminal works of Rytov \cite{Rytov:1989ur} and Polder and Van Hove \cite{Polder:1971uu}, it has been known that radiative heat transfer (RHT) can be much larger than classical Planck's prediction when the separation distance between two solid bodies are small compared to the thermal wavelength at which heat radiation is exchanged. Enhancement is particularly large in the case of dielectrics supporting phonon-polariton such as SiC or SiO$_2$ \cite{Mulet:2002we} that have probably been the most studied materials in the framework of near-field heat transfer. In the case of materials supporting polaritons, these non radiative modes can couple so that heat is transferred in vacuum through photon tunneling. However, the magnitude of the enhancement strongly depends on the amplitude and the width of the resonance in the material optical response which can be affected by its temperature \cite{BenAbdallah:2010hp,Nefzaoui:2013ia}. Indeed, when the temperature increases, one knows that phonons in general including phonon-polaritons are affected by several processes like phonon-phonon interactions so that theses modes see their lifetime decreases \cite{Ziman60}. But phonons in different materials are not equally affected by temperature changes. 

A question which naturally arises is whether the heat flux between a material at ambient temperature and another one different at a higher temperature will be the same when temperatures are inverted. If it is not the case, the two materials couple is said to exhibit thermal rectification, a thermal analog of electrical rectification better known as the diode. 

Thermal rectification has recently received a growing interest due to the emergence of thermal flow management needs related to limited energy and global warming issues.  If thermal equivalent of the diode and the transistor would be designed and fabricated, this could open the way to new passive temperature regulation and thermal circuits with no needs of electronics leading to very reliable thermal management.

In the last decade, thermal conductors exhibiting thermal rectification have been proposed. In these devices, heat carriers flux is made asymmetric by nanostructuring the material or by guiding the carriers flux differently whether they are moving in one direction or the other  \cite{Terraneo:2002tr,Li:2004vt,Li:2005vt,Chang:2006fr,Hu:2006wk,Hu:2009vg,Yang:2007ua,Segal:2008ws,Yang:2009wt}. These physical ideas can be used to conceive and fabricate a thermal transistor leading to the possibility of logical thermal circuits \cite{Wang:2007uf,Lo:2008tc}. More recently, radiative thermal rectifiers have been proposed \cite{Otey:2010if,Basu:2011ur,BenAbdallah:2013jp,Nefzaoui:2014ec,Nefzaoui:2014hw} as well as radiative thermal transistors \cite{BenAbdallah:2014jk,Joulain:2015er}. Very promising devices have been imagined based on materials exhibiting phase change transition such as Vanadium dioxyde (VO$_2$) \cite{BenAbdallah:2013jp,BenAbdallah:2014jk,vanZwol:2012ti}, superconductors  \cite{Nefzaoui:2014hw} or thermochromic materials \cite{Huang:2013vw}. The goal of this article is to show that radiative thermal rectification can also be achieved with the most studied polaritonic supported materials like SiC and SiO$_2$.

We first present optical properties measurements showing how both SiC and SiO$_2$ dielectric functions change with temperature. These measurements performed using Fourier Transform Emission Spectroscopy allow us to make near-field radiative heat transfer calculations taking into account the actual material dependance with temperature. Similar near-field calculation taking into account the temperature dependance of optical properties have already been taken in the past \cite{iizuka2012} with measurement on 3-C SiC of \cite{Olego82} but in a different situation between SiC and metallic coated-SiC. Here, we calculate heat transfer between two planar interfaces of SiC and SiO$_2$. We show that these two materials exhibit almost no thermal rectification when the exchange occurs in the far-field i.e. when the main contribution to the flux comes from the propagative waves. On the contrary, at distances below 100 nm, we show that rectification ratio drastically increases in order to reach values as high as the highest predicted and measured ones in the literature. We finally show that rectification is mainly due to the SiC resonance attenuation at high temperature.

\section{SiC and SiO$_2$ optical datas}
\begin{figure}
\begin{center}
\includegraphics[width=12cm]{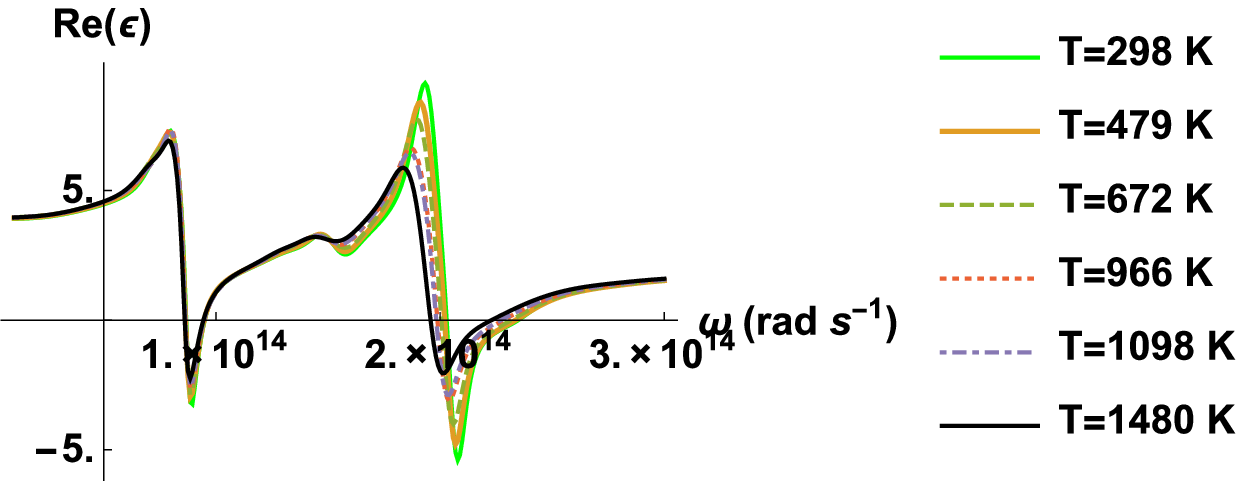}
\includegraphics[width=12cm]{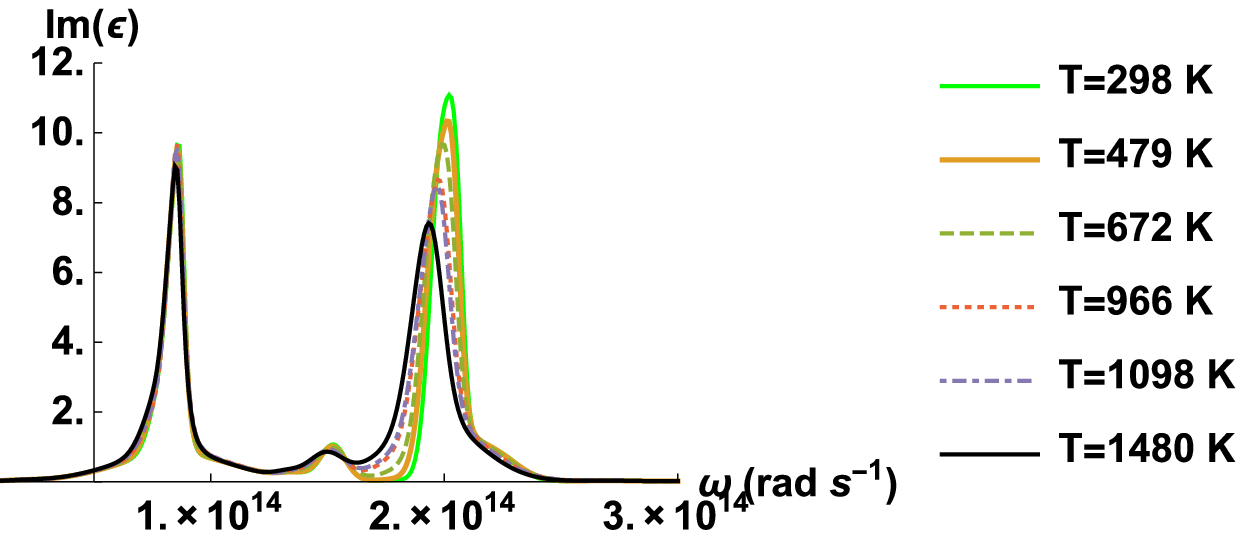}
\includegraphics[width=12cm]{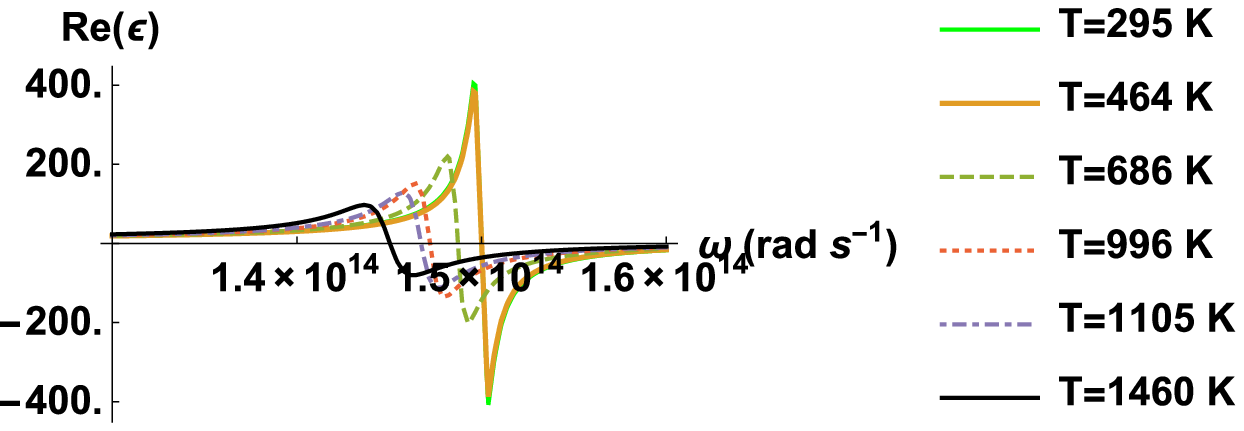}
\includegraphics[width=12cm]{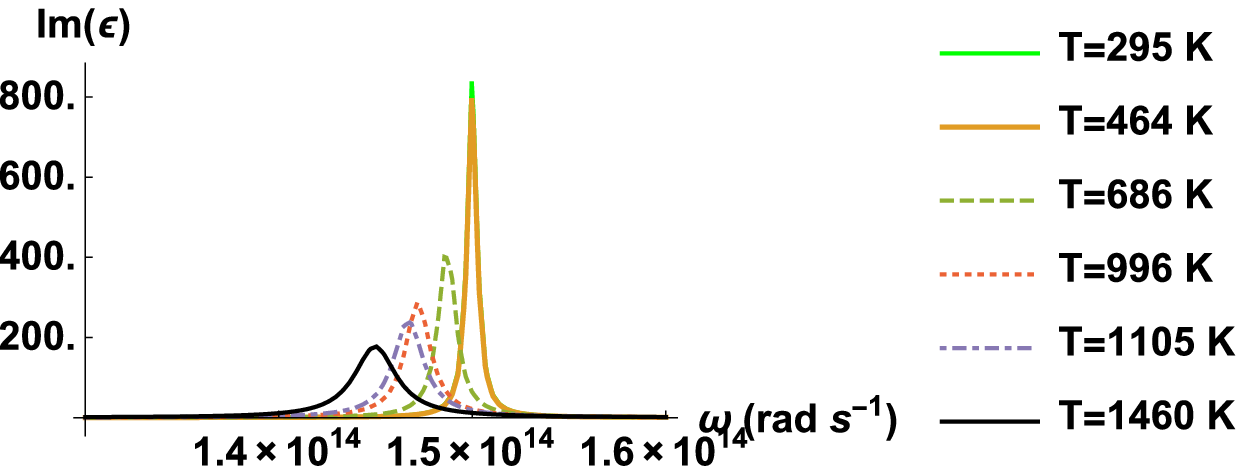}
\caption{ Measured real and Imaginary part of the dielectric function $\epsilon$ variation with angular frequency $\omega$ for different temperatures. Top 2 figures : SiO$_2$. Bottom 2 figures : SiC.}
\label{Datas}
\end{center}
\end{figure}

In this section, we report SiC and SiO$_2$ optical properties temperature dependance with temperature. The dielectric functions of SiC and vitreous SiO$_2$ have been extracted from measurements of their spectral emittance at near normal incidence. The SiC sample provided by MTI Corporation is a single crystal with a (0001) orientation and two epi polished sides. The stacking sequence is of 6H type and its electric resistivity lies between 0.02 and 0.2 ohm.cm. The SiO$_2$ material is a polished plate of vitreous silica with low OH content ($< 20 ppm$). The emittance spectra were acquired with a homemade spectrometer built around two Bruker infrared spectrometers, a Vertex 80v working under vacuum and a Vertex 70 purged with dry air (see \cite{DeSousaMeneses:2015iq}). The first and second devices have been used to probe the far- and mid-infrared ranges respectively, at a spectral resolution of 4 cm$^{-1}$. In order to avoid parasite fluxes during the measurements, a CO$_2$ laser has been used to heat the samples up to 1500 K. The retrieval of the optical functions has been performed in a second step by fitting the experimental data with adequate physical expressions. A semi-quantum dielectric function model, well adapted for crystalline media, was selected to reproduce the infrared response of the silicon carbide sample \cite{DeSousaMeneses:2004wl}. The disordered nature of vitreous silica needed instead the use of a causal Gaussian dielectric function model able to take account for the inhomogeneous broadening of the absorption bands of glasses \cite{DeSousaMeneses:2006by}. The whole process allowed to obtain the temperature dependences of the optical properties for both materials in a wide range, the results are reported in Fig. \ref{Datas}. For SiO$_2$, we note that the resonance around 2$\times$10$^{14}$ rad s$^{-1}$ slightly shifts and broadens when the temperature increases. For SiC, we note a similar behavior (shift and broadening) but even more pronounced  for the resonance around 1.5$\times$10$^{14}$ rad s$^{-1}$. This is due to the fact that anharmonicity effects responsible of the phonon-phonon interaction increases with the temperature \cite{Ziman60}. We will see that this behavior is responsible for the important rectification between SiO$_2$ and SiC in the near-field. Note however that the second resonance in SiO$_2$ does not change a lot with temperature as long as temperature is below the vitreous transition which occur around 1500 K \cite{DeSousaMeneses:2014bl}. This is probably related to the very weak SiO$_2$ thermal dilatation.

\section{Radiative heat transfer between SiC and SiO$_2$}
We now come to the calculation of the RHT between two planar interfaces made of SiC (medium 1) and SiO$_2$ (medium2). RHT between two planar interfaces can easily be calculated using fluctuational electrodynamics formalism. Heat flux appears as a semi-analytical expressions which is nothing else than the summation of individual plane wave contributions to the heat transfer. These plane waves are labelled by a triplet $(\omega,K,\alpha)$ where $\omega$ is the angular frequency of the wave, $K$ is the wavector component parallel to the interfaces whereas $\alpha$ is the wave polarization ($s$ or $p$). Let us call $T_1$ temperature of medium 1 and $T_2$ temperature of medium 2. Heat flux expression between media 1 and 2 reads \cite{BenAbdallah:2010hp,Biehs:2010kb}
\begin{equation}
\label{exprflux}
\varphi_{1\leftrightarrow 2}=\sum_{\alpha=s,p}\int_0^\infty[\Theta(\omega,T_1)-\Theta(\omega,T_2)]d\omega\int_0^\infty \frac{KdK}{4\pi^2}\tau^\alpha(\omega,K)
\end{equation}
where $\Theta(\omega,T)=\hbar\omega/[\exp[\hbar\omega/k_bT]-1]$ is the mean energy of a photon at temperature $T$.
In the preceding expression, $\tau^\alpha(\omega,K)$ appears as a transmission coefficient for a plane wave $(\omega,K,\alpha)$ between medium 1 and medium 2 or medium 2 and medium 1. Note that expressions of transmission coefficient are different whether the wave is propagative ($K<\omega/c$ or evanescent ($K>\omega/c$). If we call medium 3, the medium separating the two other media and introducing Fresnel reflection coefficients, the expression for the transmission coefficient reads for propagating waves
\begin{equation}
\label{ }
\tau^\alpha(\omega,K)=\frac{(1-|r^\alpha_{31}|^2)(1-|r^\alpha_{32}|^2)}{|1-r^\alpha_{31}r^{\alpha}_{32}e^{2i\gamma_3 d}|^2}
\end{equation}
and for evanescent waves
\begin{equation}
\label{ }
\tau^\alpha(\omega,K)=\frac{4\Im(r^\alpha_{31})\Im(r^\alpha_{32})e^{-2\Im(\gamma_3)d}}{|1-r^\alpha_{31}r^{\alpha}_{32}e^{-2\Im(\gamma_3) d}|^2}
\end{equation}
In these expressions, $\Im$ denotes the imaginary part, $\gamma_i$ is the perpendicular wave vector component in medium $i$ which for non magnetic materials reads $\gamma_i=(\epsilon_ik_0^2-K^2)^{1/2}$. The usual fresnel reflection coefficient are given by $r^s_{ij}=(\gamma_i-\gamma_j)/(\gamma_i+\gamma_j)$ and $r^p_{ij}=(\epsilon_j\gamma_i-\epsilon_i\gamma_j)/(\epsilon_j\gamma_i+\epsilon_i\gamma_j)$ when one considers local materials which is the case at nanometric separation distance \cite{Chapuis:2008kca,Singer:2015ey}. The flux expression only depends on the material temperatures, their local optical response and the separation distance between the bodies. Note that the symmetry of Eq. (\ref{exprflux}) implies that rectification will occur if and only if the two materials are different and have their optical properties that change with temperature in a different manner.

Optical properties variations with temperature have rarely been taken into account, mostly in the case of phase change materials. Here, our main goal is to perform these calculations in the case of two very well studied materials but for which optical properties variations with temperature are taken into account. The input properties in our calculation are the optical properties presented in the preceding section. It can be noted in these measurements that they have been taken at close temperatures for both materials but not exactly at the same one. As explained in the introduction, our goal is also to explore how a system constituted of SiC and SiO$_2$ can exhibit radiative thermal rectification. Thus, a difficulty arises since rectification compares a situation where the materials are at two different temperatures with a second situation where the temperatures considered before have been inverted. For example, measurements at ambient temperature for SiC have been performed at 295 K whereas they have been performed at 298 K for SiO$_2$. In our calculation, we choosed ambient temperature to be 297 K and we assumed that optical properties at this temperature were the ones at 295 K for SiC and 298K for SiO$_2$. Other temperatures considered in the paper are 471 K, 671 K, 981 K, 1102 K and 1470 K. To perform heat flux calculations, optical properties for SiC and SiO$_2$ have been respectively taken at 464 K and 479 K, 686 K and 672 K, 996 K and 966 K, 1105 K and 1098 K, 1460 K and 1480 K (See Fig. \ref{Datas}).

\begin{figure}
\begin{center}
\includegraphics[width=12cm]{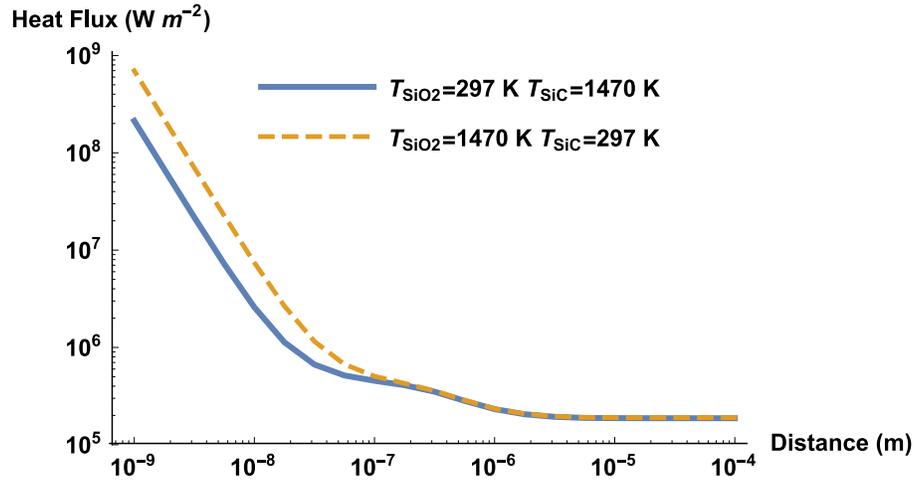}
\caption{Computed radiative heat transfer between a plane interface of SiC and a second one constituted of SiO$_2$ versus their separation distances. Two situations are compared. In the first one (plain) SiO$_2$ is at 297 K and SiC at 1470 K  whereas in the second one (dashed) SiC is at 297 K and SiO$_2$ is at 1470 K.}
\label{fl1470}
\end{center}
\end{figure}

We report in Fig. \ref{fl1470} heat flux variations as a function of the separation distance between the two materials. We compare the situation where SiO$_2$ is at 297 K and SiC at 1470 K with the situation where the temperatures are inverted. We note that for separation distances larger than a few microns the two fluxes are constant and almost superimposed each other. When SiO$_2$ is at ambient temperature, the value of the heat flux is 1.872$\times$10$^5$ W m$^{-2}$ whereas it is 1.896$\times$10$^5$ W m$^{-2}$ in the reverse case. These two values are around 71\% of the heat exchanged between two blackbodies for which Fresnel reflection factors are equal to 0 in the fluxes expression. When the separation distance is decreased, the heat flux increases for both situations. For distances larger than 200 nm, heat fluxes are very close in both situations. For distances smaller than 200 nm, we see that the situation where SiC remains at ambient temperature exhibits a higher flux than the one corresponding to the reverse one.

\begin{figure}
\begin{center}
\includegraphics[width=12cm]{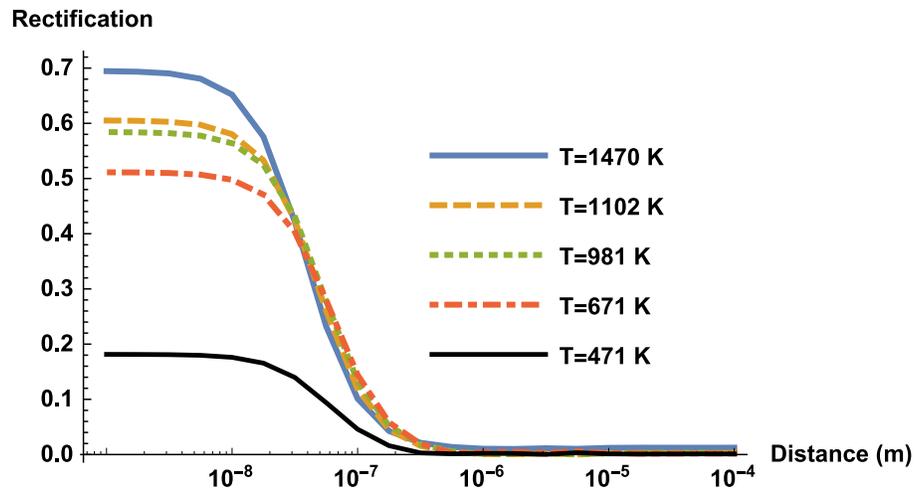}
\caption{Computed rectification variations as a function of the separation distance between two planar interfaces made of SiC and SiO$_2$ when one material is at 297 K  and the second one at 471 K, 671 K, 981 K, 1102 K or 1470 K.}
\label{Rectifvsd}
\end{center}
\end{figure}

We define rectification $R$ as the relative variation of the heat flux \cite{BenAbdallah:2013jp,Nefzaoui:2014ec} in the two situations so that
\begin{equation}
\label{ }
R=\frac{|\varphi_{1\leftrightarrow2}(T_1,T_2)-\varphi_{1\leftrightarrow2}(T_2,T_1)|}{Max[\varphi_{1\leftrightarrow2}(T_1,T_2),\varphi_{1\leftrightarrow2}(T_2,T_1)]}
\end{equation}
With such a definition, $R=0$ means the the material exhibit no rectification whereas $R=1$ correspond to the case of a perfect thermal diode. In Fig. \ref{Rectifvsd}, we report the variation of the rectification as a function of the separation distance between the two planar interfaces made of SiC and SiO$_2$ for different temperatures. One material is always at 297 K whereas the second one is taken at 471 K, 671 K, 981 K, 1102 K or 1470 K. We note a very similar behavior for all the cases presented. For distances larger than 200 nm rectification remains very low. It goes from 10$^{-3}$ for the smallest temperature difference to 10$^{-2}$ for the largest. One can therefore say that SiC and SiO$_2$ exhibit very low radiative thermal rectification in the far field and more generally for separation distances larger than 200 nm. Below 200 nm, rectification increases drastically  to reach values up to 0.7 for a temperature of 1470 K. Rectification saturates when the separation distance goes below 10 nm. Note also that the higher temperature difference is, the higher is the rectification in the near-field.

\section{Discussion}

\begin{figure}
\begin{center}
\includegraphics[width=8.3cm]{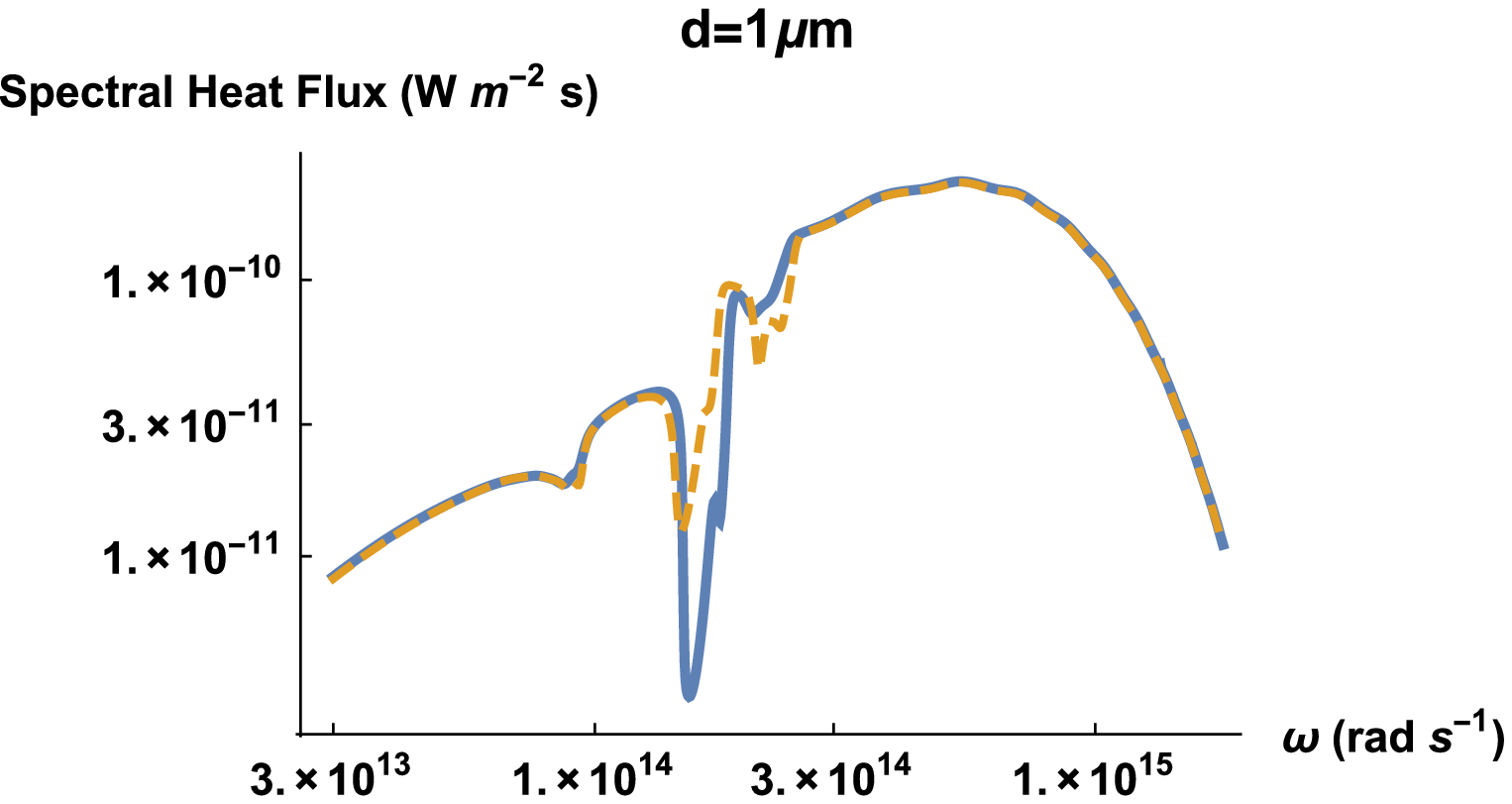}
\includegraphics[width=8.3cm]{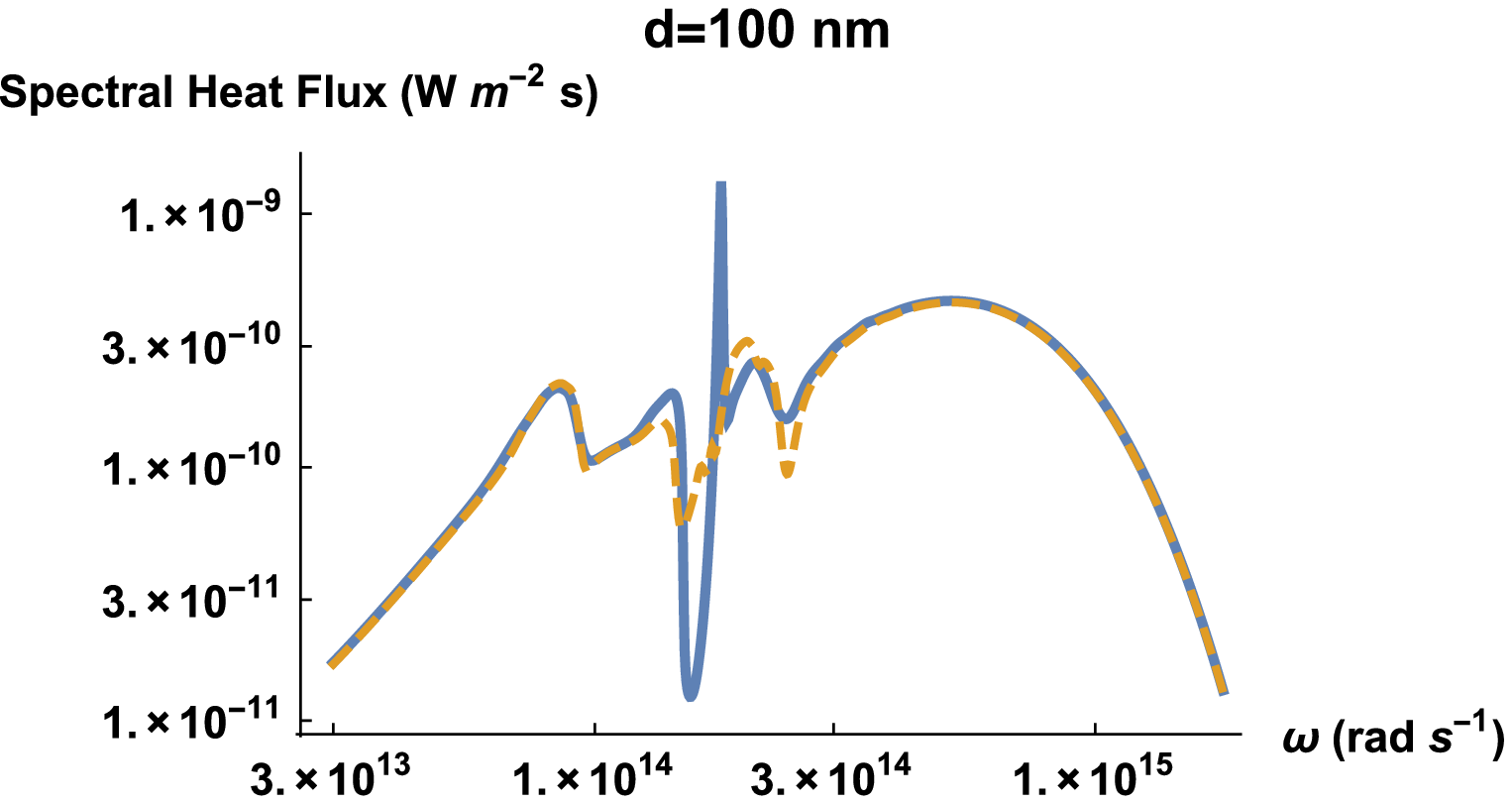}
\includegraphics[width=8.3cm]{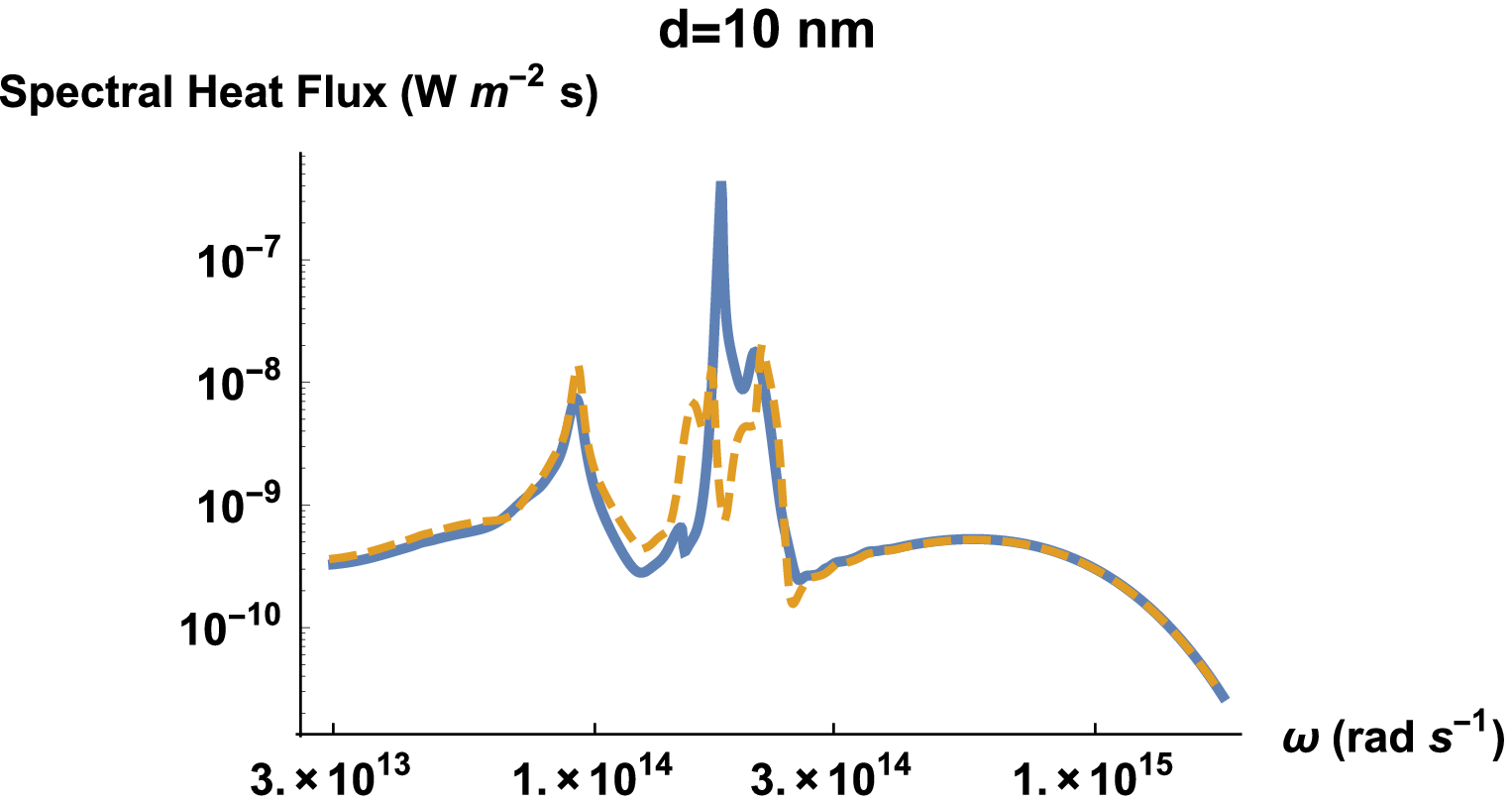}
\includegraphics[width=8.3cm]{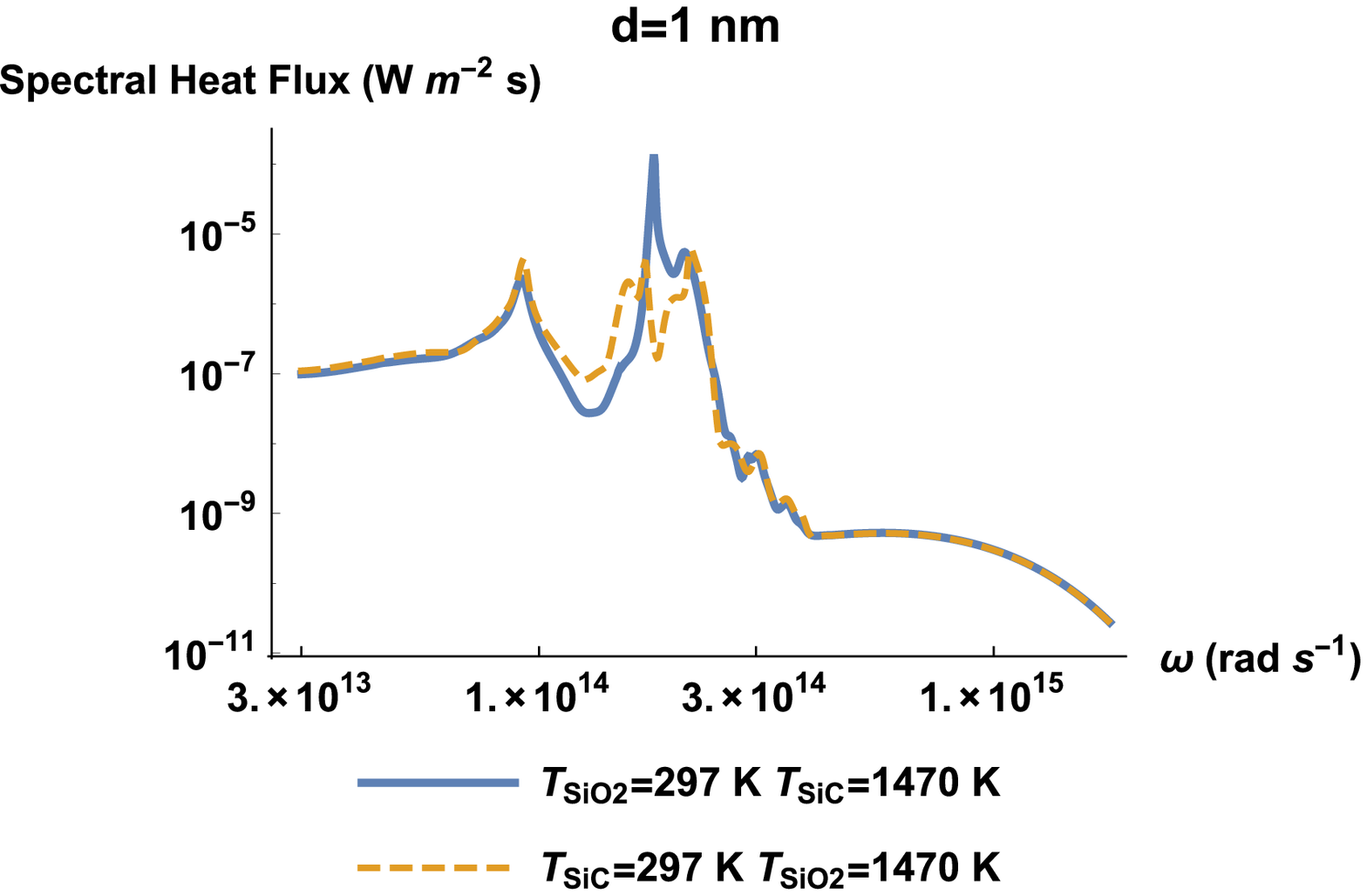}
\caption{Computed spectral radiative heat transfer between two planar interfaces made of SiC and SiO$_2$ in the direct and the reverse situation for 4 different separation distances: SiO$_2$ is at 297 K and SiC is at 1470 K in the first situation (plain) whereas SiO$_2$ is at 1470 K and SiC is at 297 K in the second situation (dashed). }
\label{specflux}
\end{center}
\end{figure}

Let us now shed some light on the physical reason that is responsible of the behavior of the rectification between SiC and SiO$_2$ with both distance and temperature. To do so, we plot in Fig. \ref{specflux} the variation of the spectral radiative heat transfer flux as a function of the angular frequency for the case where one material is at 297 K and the other one at 1470 K. The spectral flux is represented for 4 different separation distances 1 $\mu$m, 100 nm, 10 nm and 1nm.  
At large distance (here 1 $\mu$m), the spectral flux is very similar to the one representative of the exchange between two black bodies except that there exist some frequencies where one can identify dips. These dips correspond to the frequencies where SiO$_2$ and SiC exhibit resonances. Indeed, SiO$_2$ exhibit two resonances and SiC one resonance that had been identified in the first section. At these frequencies SiO$_2$ and SiC are very reflective so that transfer is greatly reduced. However, the range of frequencies where materials are low absorbent is much smaller than the one where both materials almost behave as blackbodies. This explains why at large distance when this spectral flux is integrated over all frequencies, heat flux reaches 70\% of the one between two blackbodies. If we examine carefully the region where the spectral flux is much lower than between blackbodies, one see that these are the regions where the spectral fluxes differ in the direct and the reverse situations. We notice that the depth of the dips is less pronounced when the material concerned is at high temperature. For example dips for SiO$_2$ (around 9$\times$10$^{13}$ rad s$^{-1}$ and 2$\times$ 10$^{14}$ rad s$^{-1}$) are less pronounced in the plain curve where SiO$_2$ is at high temperature. In the same way, the dip corresponding to the SiC resonance around 1.5$\times$ 10$^{14}$ rad s$^{-1}$ is smaller in the dashed curve corresponding to the case where SiC is at high temperature. Despite of these differences, as most of the contribution comes from the region where the spectral fluxes are equal, rectification in the far-field is very small and even smaller at lower temperature when the dielectric function is still close to the one at ambient temperature. 

When the distance is reduced, the situation drastically changes. The main contributions to the spectral fluxes mainly comes from the frequencies where the resonances are present. This is a well known behavior when one is treating materials supporting phonon polaritons such as SiC and SiO$_2$. Indeed, the heat flux is completely dominated by the contribution of evanescent waves in $p$ polarization \cite{Joulain:2005ih,Volokitin:2007el} which depends on the imaginary part of the static reflection coefficient which reads for medium $i$ $(\epsilon_i-1)/(\epsilon+1)$. The frequency where $\epsilon_i$ is approaching -1 corresponds to the one where there is the largest density of states of surface polaritons close to the material. 
What appears here is that the contribution of the polaritons of a material is reduced when the material is at high temperature. Moreover, one notes that SiO$_2$ and SiC do not behave in the same way. While the SiO$_2$ resonance contribution is weakly attenuated, the one of SiC is much more decreased by almost one order of magnitude above 1000 K. This explains why the heat flux is larger when the SiC is at low temperature. Moreover rectification does not change below a certain distance which correspond to the one where the flux is completely dominated by the $p$ contribution of the evanescent waves. It has been shown in several papers  \cite{Mulet:2002we,Joulain:2005ih,Rousseau:2012bb}, that this flux behaves as $1/d^2$ multiplied by the product of the imaginary parts of the reflection coefficients. When the distance decreases, both fluxes increase following the same distance scaling law. Therefore, rectification becomes distance independent and saturates.

\section{Conclusion}
We have presented in this paper calculation of the radiative heat flux between two materials SiC and SiO$_2$ taking into account the real temperature variations of their optical properties. To do so, we have used measurements of the optical response of both SiO$_2$ and SiC at different temperatures from the ambient one up to almost 1500K. We show that when the temperature difference between the two samples is large, there is little difference in the far field between the fluxes calculated in a situation where one of the materials is at 297 K and the other one at higher temperature and another situation where the temperatures are reversed. We see that at any temperature difference, rectification increases between 200 nm and 10 nm to reach a value which is crescent with the temperature difference. Using the data measured here, one has shown that this rectification can reach values up to 0.7, values that are as high as the one obtained with phase change materials \cite{BenAbdallah:2013jp,Ito:2014gf}. These properties could be exploited in thermal management. Thermal radiative rectification could for example be adjusted by controlling the separation distance between the materials  and their temperatures.

\section*{Acknowledgments}
This work has been supported by CNRS through the grant CARDYMOL of the ``Cellule \'energie''. This work also pertains to the French Government Program ``Investissement d'Avenir" (LABEX INTERACTIFS, ANR-11-LABX-0017-01).

\end{document}